\documentclass[letterpaper, 10 pt, conference]{ieeeconf}  
\IEEEoverridecommandlockouts                                                        \usepackage{epsfig} 
\usepackage{graphicx}
\usepackage{amsmath}
\usepackage{array}
\usepackage{amssymb}
\usepackage{verbatim}
\usepackage{cleveref}
\usepackage{color}
\usepackage{relsize}
\usepackage{tikz}
\usepackage{fancyhdr}
\usepackage{graphics}
\usepackage{setspace}
\usepackage{epstopdf}
\usepackage{multirow}
\usepackage{lipsum}
\usepackage{textcomp}
\usepackage{mathtools}
\usepackage{cite}
\usepackage{setspace}
\usepackage{booktabs}
\usepackage{breqn}
\usepackage{algorithm}
\usepackage[noend]{algpseudocode}
\usepackage{flushend}
\usepackage{multicol}
\usepackage{float}
\usepackage{mathtools,bm}
\usepackage[noend]{algpseudocode}
\usepackage{booktabs}
\usepackage{dblfloatfix}
\usepackage{mathrsfs}
\usepackage{mathtools}

\fancypagestyle{firstpage}{\lhead{\vspace{-0.75 cm} \small \textbf{{\fontfamily{cmss}\selectfont
2020 American Control Conference (ACC)\\July 1--3, 2020, Denver, CO, USA}}}}

\title{\Large \bf
Integrated Power and Thermal Management of Connected HEVs\\ via Multi-Horizon MPC$^*$}

\author{Qiuhao Hu$^{1}$, Mohammad Reza Amini$^{1}$, Hao Wang$^{1}$, Ilya Kolmanovsky$^{2}$, and Jing Sun$^{1}$%
\thanks{*This paper is based upon the work supported by the United States Department of Energy (DOE) under award No. DE-AR0000797.}
\thanks{$^{1}$Q. Hu, M.R. Amini, H. Wang and J. Sun are with the Department of Naval Architecture \& Marine Engineering, University of Michigan, Ann Arbor, MI 48109 USA. 
Emails: {\tt\small \{qhhu,mamini,autowang,jingsun\}@umich.edu}}
\thanks{$^{2}$I. Kolmanovsky is with  the Department of Aerospace Engineering, University of Michigan, Ann Arbor, MI 48109 USA. Email: {\tt\small ilya@umich.edu}}%
}

\renewcommand{\baselinestretch}{0.95}
\begin{document}

\maketitle
\thispagestyle{firstpage}

\begin{abstract}
In this paper, a multi-horizon model predictive controller ({MH}-MPC) is developed for integrated power and thermal management (\textit{iPTM}) of a power-split hybrid electric vehicle (HEV). The proposed \text{MH}-MPC leverages an accurate short-horizon vehicle speed preview and an approximate forecast over a longer shrinking horizon till the end of the driving cycle. This multiple-horizon scheme is developed to cope with fast and slow dynamics associated with power and thermal responses. The main objective of the proposed \text{MH}-MPC is to minimize fuel consumption and enforce the power and thermal constraints on the battery state-of-charge and engine coolant temperature, while meeting the driving (traction) and cabin air conditioning (heating) demands. The proposed \text{MH}-MPC allows for exploiting the engine coolant as thermal energy storage, providing more flexibility for the HEV energy flow optimization. The simulation results show that the proposed \text{MH}-MPC provides near-optimal results in reference to the Dynamic Programming (DP) solution with an affordable computational cost. Moreover, compared with a more conventional MPC strategy, the \text{MH}-MPC can leverage the speed previews with different resolutions effectively to achieve the desired performance with satisfactory robustness.
\end{abstract}

\vspace{-0.1cm}
\section{INTRODUCTION}
Efficient thermal management of hybrid electrical vehicles (HEVs), including engine cooling, cabin heating/cooling, and aftertreatment system, has a significant impact on the overall vehicle energy efficiency~\cite{gong2019integrated,kim2016thermal,Amini_CCTA19,kim2014thermal}. Power and thermal loops of an HEV are strongly coupled. For example, the engine coolant temperature has a direct impact on engine efficiency, emission, cabin air conditioner (i.e., cabin heating in winter). While numerous energy management strategies (EMS) have been developed for HEVs with the focus on power-split and energy flow management to meet the traction power demand and improve fuel economy~\cite{malikopoulos2014supervisory,tie2013review,zhang2015comprehensive},~the integrated power and thermal management (iPTM) of HEVs and plug-in HEVs (p-HEVs) has been the subject of only a few recent studies~\cite{wei2019optimal,shahbakhti2019,shams2012integrated}. 

EMS strategies are based on either heuristic or optimization methods. Many optimization-based approaches, such as dynamic programming (DP)~\cite{lin2003power,brahma2000opt,amini2019sequential} and Pontryagin’s maximum principle (PMP)~\cite{musardo2005ecms,kim2011optimal}, have been applied under the assumption that the full driving cycle is known a \textit{priori}. DP with multiple states is computationally demanding, making its real-time implementation infeasible. Compared with DP, PMP reduces the computational time and can be implemented online with real-time adaptation, i.e., adaptive PMP~\cite{musardo2005ecms,onori2011adaptive}. A common assumption considered in the PMP is to set the co-state of the battery state-of-charge ($SOC$) as a constant, thereby simplifying the adaptation processes. However, for the iPTM problem, the thermal states (e.g., engine coolant temperature) introduce additional co-states, which usually cannot be regarded as constants. The increased complexity
due to multiple co-states makes it harder for real-time adaptation. Another limitation of the PMP-based methods is the difficulty in handling state constraints.

Model predictive control (MPC) is an online optimization-based technique extensively investigated in previous studies~\cite{borhan2011mpc,zhang2016real,wang2016model} for energy management of HEVs. While MPC can handle the state and input constraints explicitly and is computationally less demanding, as compared to DP, it strongly relies on the prediction of the future vehicle speed. Uncertainties in the speed prediction can raise robustness challenges. With the recent development in the connectivity-based technologies, e.g., vehicle-to-vehicle (V2V) and vehicle-to-infrastructure (V2I) communications, short-range prediction of the future vehicle speed and traffic events has become more feasible and accurate~\cite{yang2019eco}. Long-term vehicle speed prediction, however, is still subject to large uncertainties. Thus, the MPC-based solutions are often implemented with a short prediction horizon to reduce the impact of the speed preview uncertainties and lower the computational load. This approach, on the other hand, limits the energy-saving potentials of the MPC-based EMS~\cite{Amini_TCST_2019}. This issue is more pronounced for iPTM as the thermal and power systems have different timescales and require different lengths of prediction horizon to optimize their responses~\cite{Amini_TCST_2020}.

In order to address the challenges associated with MPC-based iPTM of HEVs, we introduce a novel multi-horizon MPC (MH-MPC) framework in this paper. The main objective of the MPC-based iPTM is to minimize the fuel consumption, while enforcing the power and thermal constraints in response to the traction and cabin heating demands. The operations in cold weather conditions are considered in the thermal loops. We consider the entire time horizon of the trip and divide it into two segments. The first segment is relatively short with high resolution, during which the vehicle speed predictions could be obtained from the V2I/V2V information and is assumed to be accurate. The second segment is a shrinking horizon with low resolution to reduce the computation demands of the predictive controller. An approximate long-term prediction of the future vehicle speed, which can be realized by processing the traffic data collected from the connected vehicles~\cite{Amini_TRB20}, is incorporated in the second segment of the horizon. 

The main contributions of this paper are twofold. Firstly, a multi-horizon MPC is developed to leverage the speed previews with different accuracies over short and long prediction horizons. This multi-horizon framework not only reduces the computational demands, but it also allows for incorporating fast (i.e., power) and slow (i.e., thermal) dynamics over different prediction horizons. Secondly, the engine coolant is exploited as thermal energy storage, providing further flexibility for hybrid energy flow optimization. The MH-MPC allows us to utilize the thermal energy storage to the full extent with the multi-horizon speed preview.

The rest of the paper is organized as follows: experimentally validated power and thermal models of a power-split HEV are described in Sec.~\ref{sec:sec_2}. Next, the baseline (conventional) MPC and the proposed MH-MPC are presented in Sec.~\ref{sec:sec_3}. Sec.~\ref{sec:sec_4} reports simulation results. Finally, the concluding remarks are summarized in Sec.~\ref{sec:sec_5}. 

\vspace{-0.1cm} 
\section{HEV Power and Thermal Models}
\label{sec:sec_2}
\begin{figure}[h!]
	\begin{center}
		\includegraphics[width=0.85\columnwidth]{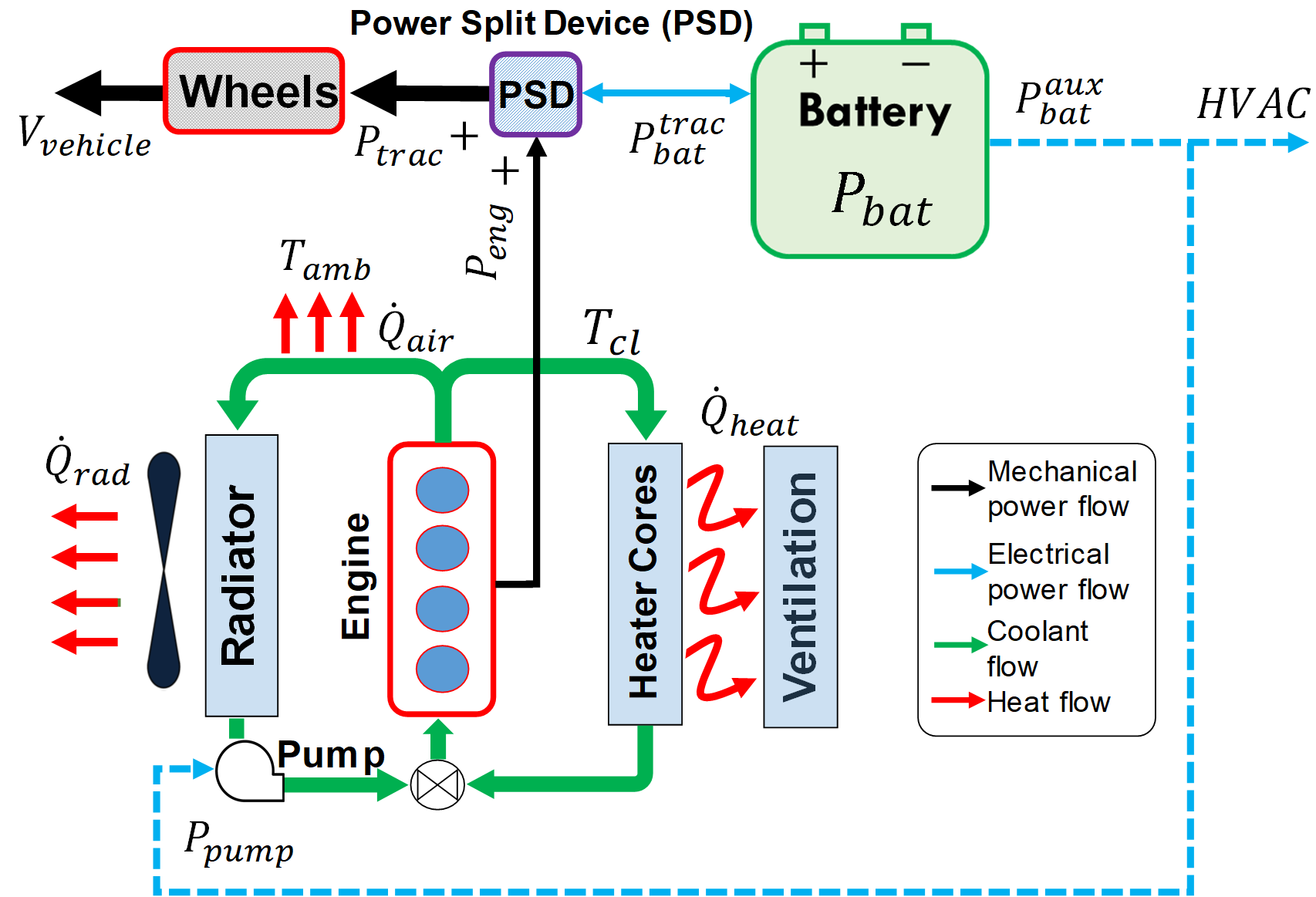}
		\vspace{-0.35cm}   
		\caption{Schematic of power-split HEV thermal and power loops.}\vspace{-0.7cm} \label{fig:HEV_PS_AC_Schematic} 
	\end{center}
\end{figure}

Consider the powertrain for the power-split Toyota Prius HEV. The overall schematic of the power and thermal loops of the HEV is shown in Fig.~\ref{fig:HEV_PS_AC_Schematic}. The two states of interest, which represent two energy storages within the HEV, are battery $SOC$ and engine coolant temperature ($T_{cl}$). We presented physics-based models of $SOC$ and $T_{cl}$ dynamics in our previous work~\cite{gong2019integrated}, where the power ($SOC$ and fuel consumption rate $\dot{m}_{fuel}$) and thermal ($T_{cl}$) models were experimentally validated against the data collected from a Prius HEV MY 2017. 
\vspace{-0.35cm}

\vspace{-0.15cm}   
\subsection{Battery Power-Balance Model} \vspace{-0.1cm}
%
The battery provides electrical traction power ($P_{bat}^{trac}$), as well as auxiliary power ($P_{bat}^{aux}$), e.g., for an air conditioning (HVAC) system. The battery $SOC$ is modeled using the equivalent circuit model~\cite{Amini_TCST_2019}:   
%
\vspace{-0.15cm}
\begin{gather}\label{eq:SOC_simple_model}
\dot{SOC}(t)=\frac{U_{oc}(t)-{\sqrt {U_{oc}^2(t)-4R_{int}(t)P_{bat}(t)} }}{{2{R_{int}(t){C_{bat}}}}},
\end{gather}
where $P_{bat}=P_{bat}^{trac}+P_{bat}^{aux}$, $C_{bat}$, $R_{int}$ and $U_{oc}$ are the battery power, capacity, internal resistance, and open-circuit voltage, respectively. Note that $P_{bat}^{trac}$ is a part of the total traction power demand ($P_{d}$) and the rest is delivered by the combustion engine:
%
\vspace{-0.25cm}
\begin{gather}\label{eq:Power_demand}
P_{d}(t)=P_{bat}^{trac}(t)+P_{eng}(t),
\end{gather}
where the engine mechanical output power ($P_{eng}$) is determined by engine speed ($\omega_{e}$) and torque ($\tau_{e}$), which follow the optimal operating points (OPP) line on which engine brake specific fuel consumption (BSFC) is minimized.
%
\vspace{-0.1cm}   
\subsection{Engine Coolant Temperature Model}
\begin{figure}[h!]
	\begin{center}
		\includegraphics[width=0.8\columnwidth]{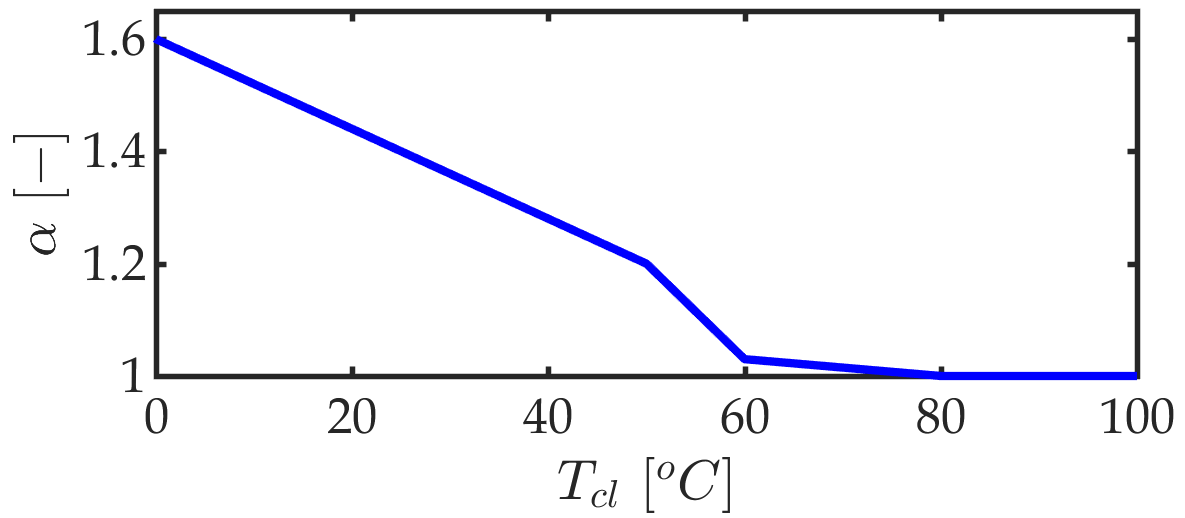}
		\vspace{-0.4cm}   
		\caption{The correction multiplier reflecting the impact of coolant temperature on the actual engine fuel consumption at low temperatures}
		\vspace{-0.6cm} \label{fig:thermal_correction} 
	\end{center}
\end{figure}

The engine thermal dynamics are represented by the coolant temperature model~\cite{gong2019integrated},
%
\vspace{-0.15cm}
\begin{gather}\label{eq:thermal_temp}
\dot{T}_{cl}=\frac{1}{M_{eng}C_{eng}}({\dot{Q}}_{fuel}-P_{eng}-{\dot{Q}}_{exh}-{\dot{Q}}_{air}-{\dot{Q}}_{heat}),
\end{gather}
where $M_{eng}$ and $C_{eng}$ are the equivalent thermal mass and capacity of the engine cooling system, respectively. Additionally, ${\dot{Q}}_{fuel}$
is the heat rate released in the combustion process, ${\dot{Q}}_{exh}$ is the heat rate rejected in the exhaust, ${\dot{Q}}_{air}$ is the rate of the heat rejected by air convection, and ${\dot{Q}}_{heat}$ is the heat rate exchanged for cabin heating. In~(\ref{eq:thermal_temp}), ${\dot{Q}}_{fuel}$ is calculated as a function of the fuel consumption rate and lower heating value ($LHV$) of gasoline. The fuel consumption rate is a function of many variables, among which we consider engine speed, engine torque, and engine coolant temperature:
\vspace{-0.15cm}
\begin{gather}\label{eq:heat_fuel}
{\dot{Q}}_{fuel}=LHV\cdot \dot{m}_{fuel}(\omega_{e},\tau_{e},T_{cl})\\
\label{eq:fuel_rate}
\dot{m}_{fuel}(\omega_{e},\tau_{e},T_{cl})=\alpha(T_{cl})\cdot f_{fuel}(\omega_{e},\tau_{e})
\end{gather}
where $\dot{m}_{fuel}$ is the fuel consumption rate, $f_{fuel}(\omega_{e},\tau_{e})$ is the nominal fuel consumption rate calculated according to the BSFC map and $\alpha(T_{cl})$ is a correction multiplier introduced to reflect the impact of $T_{cl}$. The function of $\alpha(T_{cl})$, shown in Fig.~\ref{fig:thermal_correction}, can be found in Autonomie\footnote{Autonomie\textsuperscript{\textregistered} is a MATLAB\textsuperscript{\textregistered}/Simulink\textsuperscript{\textregistered}-based system simulation tool for vehicle energy consumption and performance analysis developed by Argonne National Laboratory (ANL)~\cite{kim2014thermal}} software's thermal HEV model. 
\vspace{-0.3cm}  

\vspace{-0.25cm}
\section{MPC-based iPTM of HEV} \label{sec:sec_3}
In this paper, we consider winter time operations when cabin heating is required and the coolant temperature should be maintained within a specified range. We assume that the cabin heating demand, ${\dot{Q}}_{heat}=1.5~kW$, is constant and the engine is warmed-up at the beginning of the trip ($T_{cl,init}=50^oC$), which means the engine cold-start is not considered. In the following subsections, first, a baseline MPC with a short prediction horizon and with a quadratic term in the stage cost is formulated. Next, and based on the insight gained from the baseline MPC, a multi-horizon MPC (MH-MPC) is proposed. In both cases, battery power is considered as the control input in MPC design and the prediction model is a two-state model reflecting the dynamics of $T_{cl}$ and $SOC$.

\vspace{-0.15cm}   
\subsection{Baseline (Conventional) MPC}
The baseline MPC solves the following finite-time constrained optimal control problem:
\vspace{-0.15cm}
\begin{equation}\label{eq:baseline_MPC_formulation}
\begin{split}
\text{min}~\ell = \text{min} \{ \sum_{i=t}^{t+H-1}{\dot{m}}_{fuel}\big(P_{eng}(i),T_{cl}(i)\big)\delta t~~~~~~~\\
+\lambda\big(SOC(t+H)-SOC_{ref}\big)^2 \}
\end{split}
\end{equation}
subject to:\vspace{-0.35cm}
\begin{gather}\label{eq:baseline_constraints_1}
SOC_{min}\le SOC(i) \le SOC_{max},\\
T_{cl,min}\le T_{cl}(i) \le T_{cl,max},\label{eq:baseline_constraints_2}
\end{gather}
where $H$ is the prediction horizon, $t$ indicates the current time, $\delta t$ is the sampling time, and $i=t,\cdots,t+H$. Other parameters for $SOC$ and $T_{cl}$ in (\ref{eq:baseline_constraints_1}) and (\ref{eq:baseline_constraints_2}) are: $SOC_{min}=0.4$, $SOC_{max}=0.8$, $T_{cl,min}=40^oC$, and $T_{cl,max}=90^oC$. The cost function $\ell$ consists of two terms, (i) the integration of fuel consumption rate, and (ii) the quadratic penalty term on terminal $SOC(t+H)$. This quadratic term, with a relatively large and constant weighting factor of $\lambda$, is included to enforce the battery charge sustainability constraints. Note that here we set the reference $SOC$ as $SOC_{ref}=SOC_{init}$. A similar MPC framework has been adopted in the literature~\cite{borhan2011mpc,wang2016model,poramapojana2012minimizing}.

The optimization problem (\ref{eq:baseline_MPC_formulation}) is solved every $1~sec$, and the resulting $P_{bat}(t)$ is used to determine the required engine power according to (\ref{eq:Power_demand}), based on which the desired engine speed and torque are obtained according to the optimal BSFC map. The MPC is solved numerically using MPCTools~\cite{risbeck2016mpctools} package, which exploits CasADi~\cite{andersson2019casadi} for automatic differentiation and IPOPT algorithm for the numerical optimization. 

\begin{figure}[h!]
	\begin{center}
		\includegraphics[width=0.9\columnwidth]{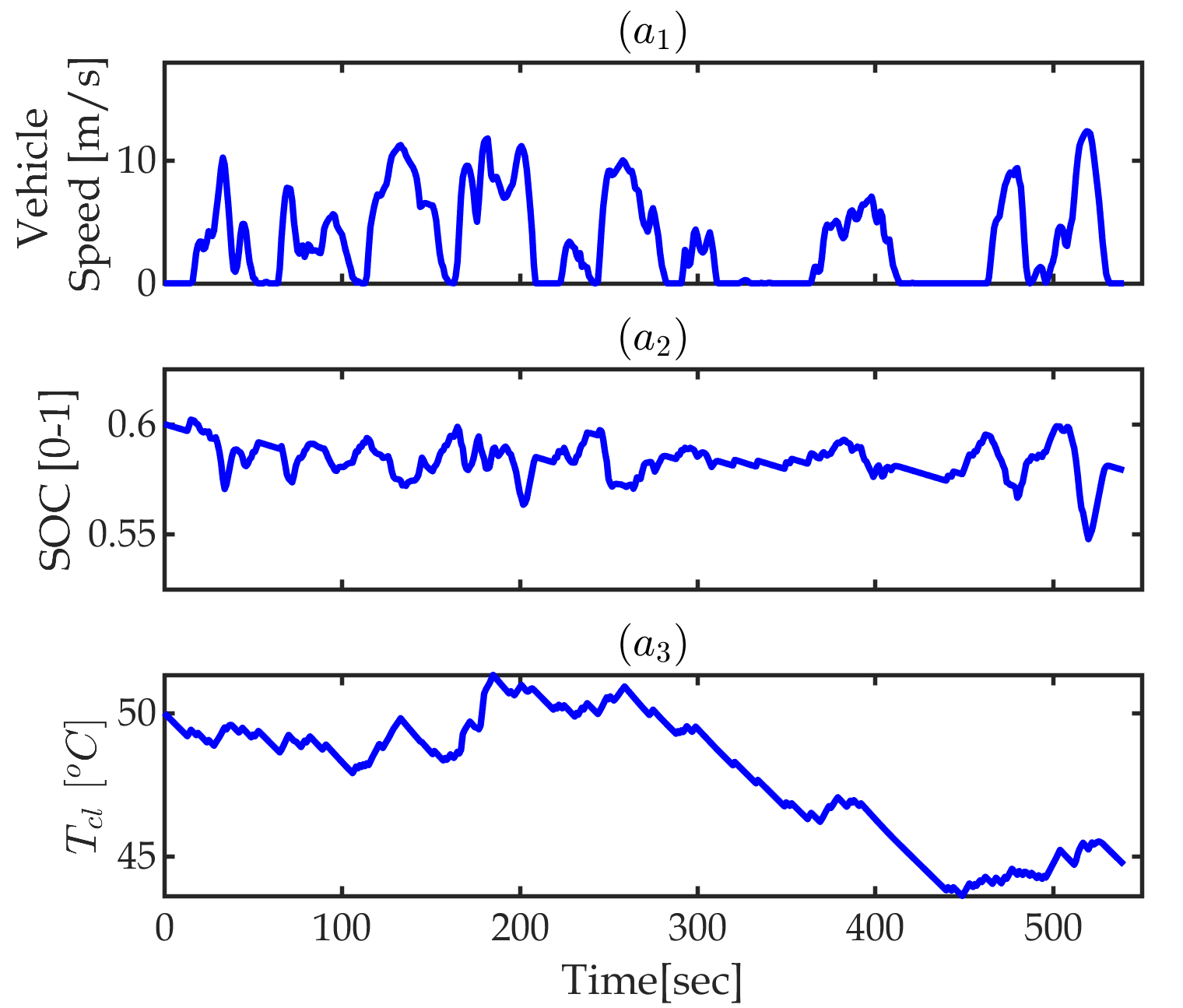}
		\includegraphics[width=0.895\columnwidth]{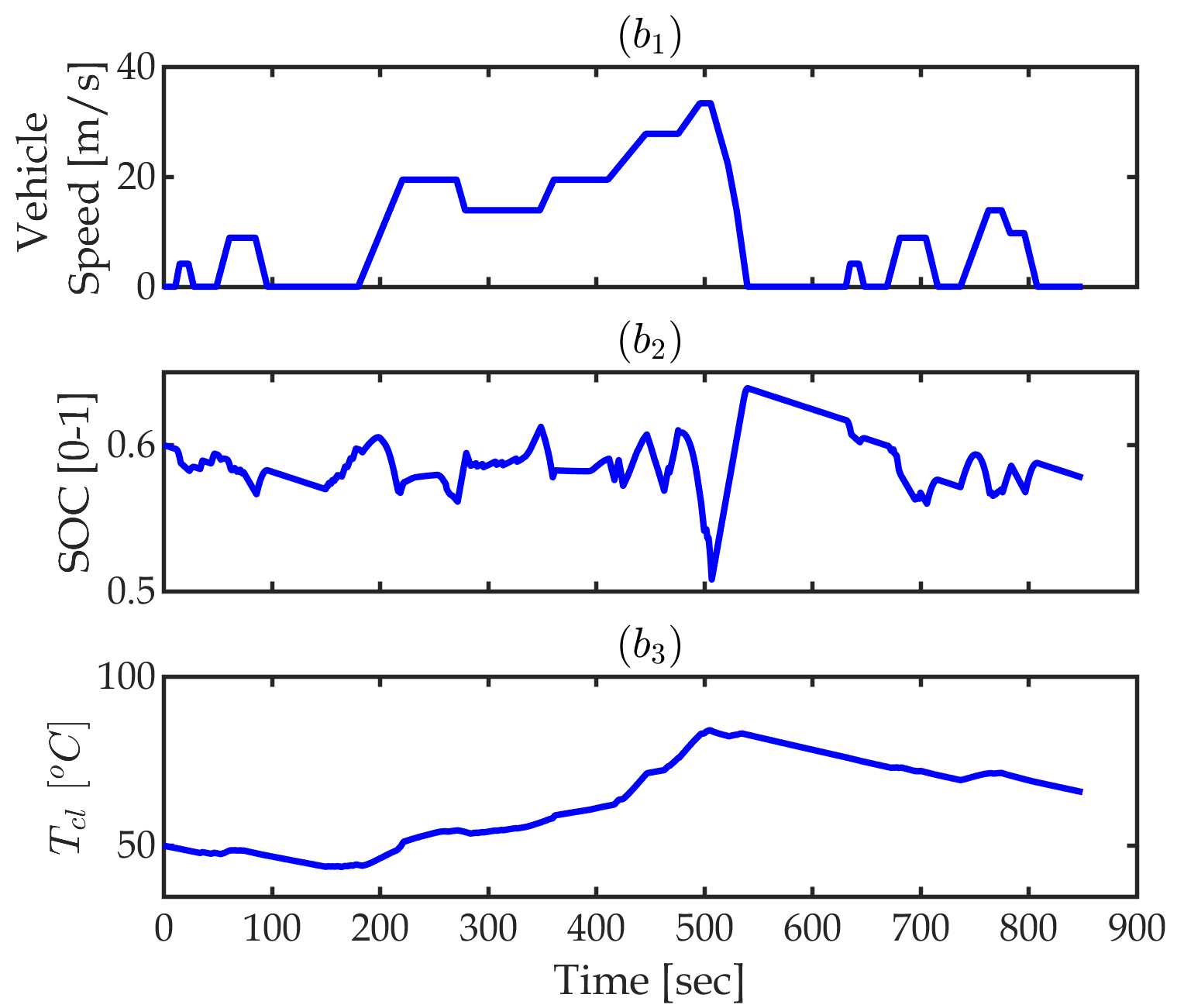}
		\vspace{-0.35cm}   
		\caption{The state
		trajectories ($SOC$, $T_{cl}$) of the baseline MPC over (a) NYCC (b) truncated NEDC.}
		\vspace{-0.65cm} \label{fig:Baseline_Results_1} 
	\end{center}
\end{figure}

The baseline MPC (\ref{eq:baseline_MPC_formulation}) is simulated over two different driving cycles, New York City Cycle (NYCC) representing urban driving conditions with multiple stop-and-go, and a truncated version of the New European Driving Cycle (NEDC). These two driving cycles are shown in Fig.~\ref{fig:Baseline_Results_1}-($a_1$) and ($b_1$), respectively. First, the MPC prediction horizon is set to $H=20~(20~sec)$. The numerical simulation results of the baseline MPC are shown in Fig.~\ref{fig:Baseline_Results_1} for both of the considered driving cycles. It can be seen in Figs.~\ref{fig:Baseline_Results_1}-($a_2$) and ($b_2$) that the battery $SOC$ is varying within a small range ($<10\%$). This is a direct result of the penalty term in the cost function (\ref{eq:baseline_MPC_formulation}) and the relatively short prediction horizon. Similar responses were also reported in~\cite{borhan2011mpc,poramapojana2012minimizing,wang2016model}.
\begin{figure}[h!]
	\begin{center}
		\includegraphics[width=1.0\columnwidth]{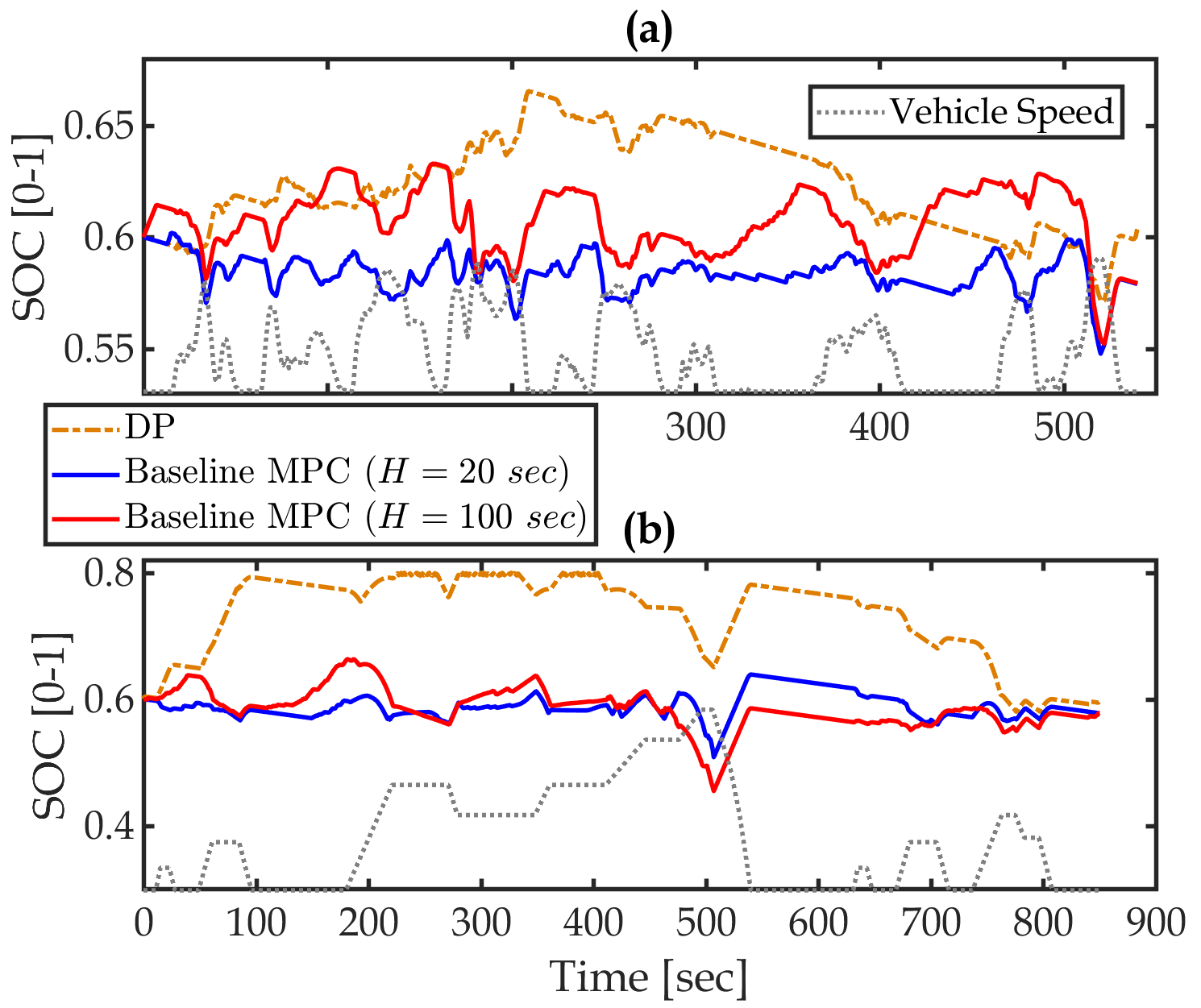}
		\vspace{-0.75cm}   
		\caption{Numerical results of $SOC$ from baseline MPC with different prediction horizon ($H$ with $\delta t_1=1~sec$) and DP over (a) NYCC (b) truncated NEDC.}
	 \label{fig:Increasing_Horizon_1} 
	\end{center}\vspace{-0.85cm}
\end{figure}

One approach to enable the controller to exploit a wider $SOC$ is by extending the prediction horizon. Fig.~\ref{fig:Increasing_Horizon_1} shows the results of simulating the same baseline MPC in (\ref{eq:baseline_MPC_formulation}) with $H=20$ and $H=100$. Moreover, the $SOC$ trajectory from the DP is plotted for both driving cycles in Fig.~\ref{fig:Increasing_Horizon_1}. It can be seen that by increasing the prediction horizon from $20$ to $100$, the variation range of $SOC$ slightly increases. By comparing the MPC results with those from the DP solution, Fig.~\ref{fig:Increasing_Horizon_1} shows that even with a longer prediction horizon, the baseline MPC delivers sub-optimal $SOC$ trajectories, i.e., the battery not being utilized efficiently.

\begin{figure}[t!]
	\begin{center}
		\includegraphics[width=0.95\columnwidth]{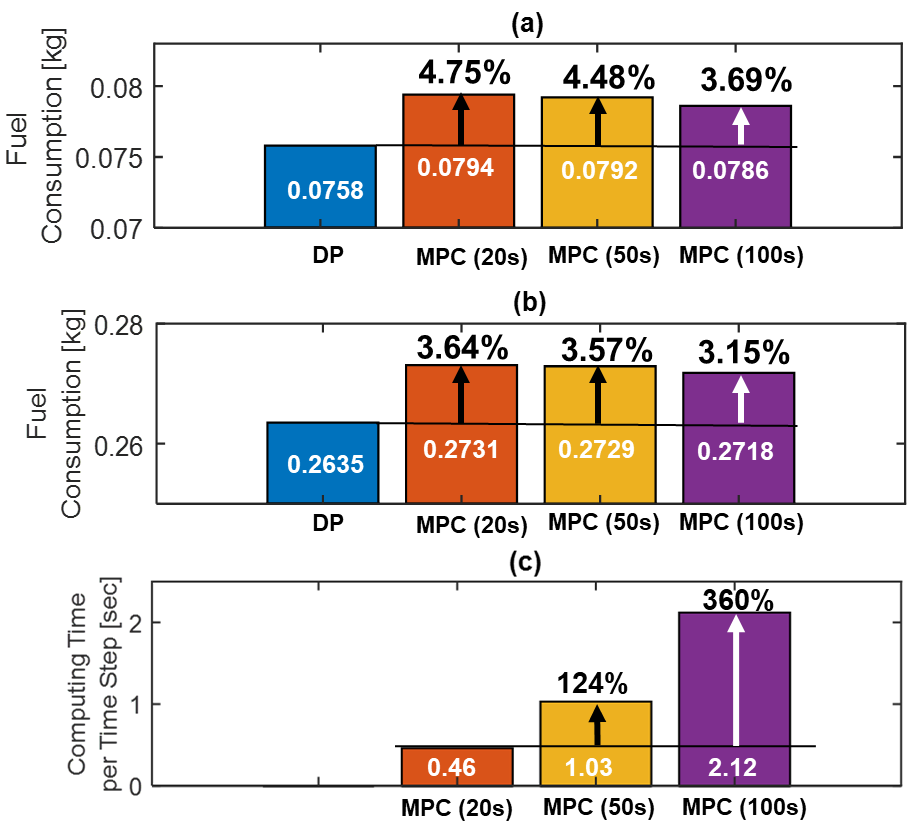}
		\vspace{-0.35cm}   
		\caption{Fuel consumption results from baseline MPC with different prediction horizon lengths ($H=20~(20~sec),~50~(50~sec),~100~(100~sec)$) and DP over (a) NYCC (b) truncated NEDC. The average MPC computation time per step time is shown in subplot (c) for different prediction horizons.}
		\vspace{-0.85cm} \label{fig:Fuel_Baseline} 
	\end{center}
\end{figure}

The fuel consumption results of the baseline MPC with different prediction horizons ($20,~50,~100$), along with the DP results are summarized in Fig.~\ref{fig:Fuel_Baseline}. While increasing the prediction horizon reduces the MPC fuel consumption, the baseline MPC consumes $3\%$ more fuel than the DP solutions. This observation shows that increasing the baseline MPC prediction horizon has a marginal impact on the fuel economy, mainly due to the impact of the $SOC$ penalty term in the MPC cost. Extending the prediction horizon imposes several major challenges in the MPC implementation. Firstly, as shown in Fig.~\ref{fig:Fuel_Baseline}-($c$), it increases the computational time of the controller significantly on an Intel\textsuperscript{\textregistered} Core i7-8700@3.20 GHz processor. Secondly, larger uncertainty associated with the long-term vehicle speed prediction can degrade the performance of the MPC~\cite{AminiCDC18,Amini_CDC19}. 

\vspace{-0.15cm}   
\subsection{Multi-Horizon MPC (MH-MPC)}
\begin{figure}[h!]
	\begin{center}
		\includegraphics[width=0.95\columnwidth]{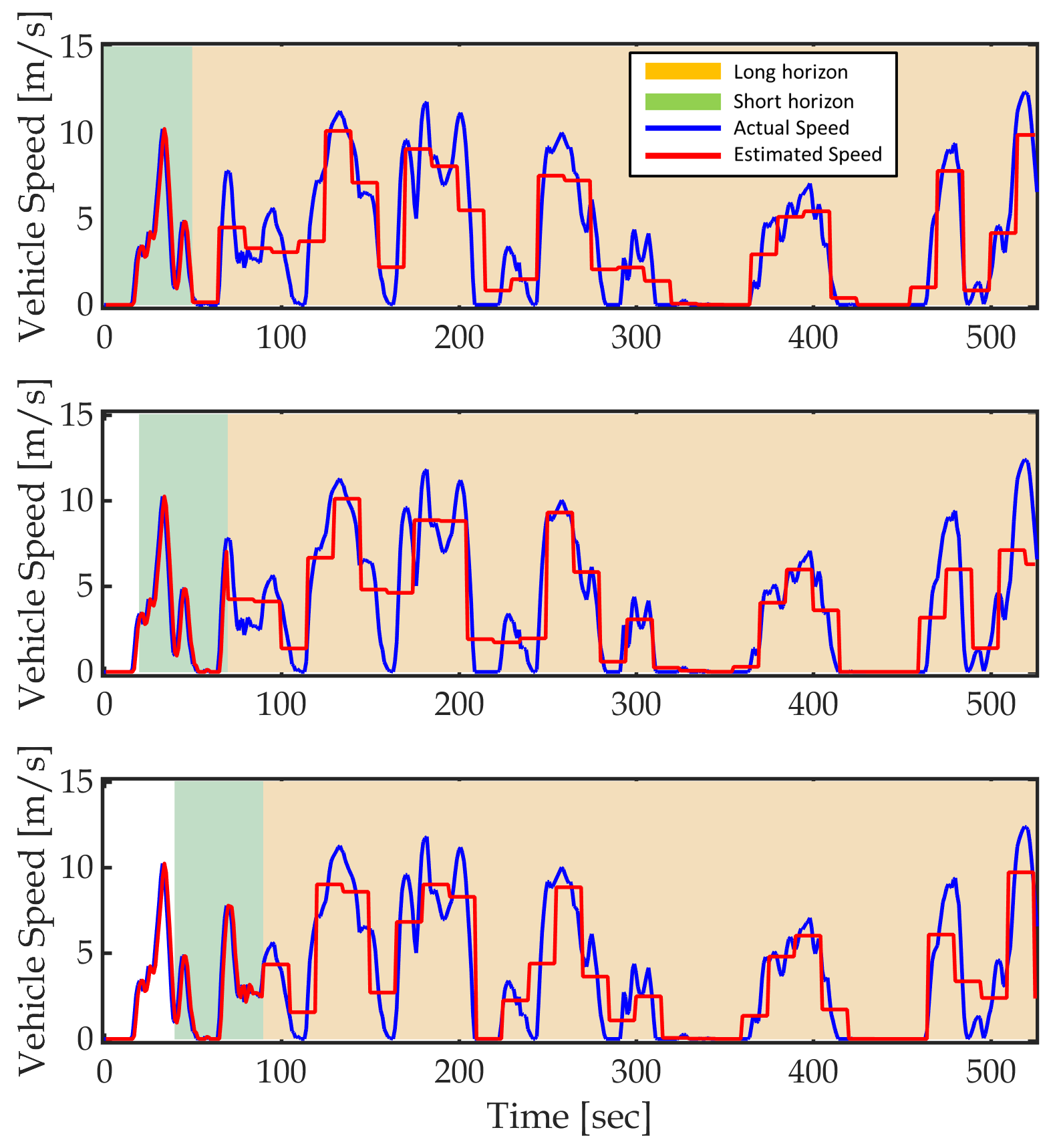}
		\vspace{-0.45cm}   
		\caption{The concept of multi-horizon incorporated in MH-MPC with short (high resolution) and long (low resolution) term vehicle speed predictions.}
		\vspace{-1.35cm} \label{fig:future_information} 
	\end{center}
\end{figure}

It was shown in the previous section that a short-horizon MPC with a penalty term in the cost function may not allow for efficient use of the vehicle speed preview. On the other hand, long-horizon MPC is significantly more computationally demanding and its performance is degraded by the uncertainties in the long-term vehicle speed predictions. To bridge this gap, a multi-horizon MPC (MH-MPC) framework is now introduced. The MH-MPC has a long-prediction horizon and assumes that the end of the trip is known a \textit{priori}. While the entire driving cycle is unknown, we assume that:
\begin{itemize}
    \item an \textbf{accurate} forecast of the vehicle speed over a short prediction horizon is available based on V2V/V2I communication, see~\cite{amini2019sequential,Amini_CCTA19}.
    \item an \textbf{approximate} forecast of the vehicle speed beyond the short-range accurate prediction window is also available. Such a forecast can be generated by processing traffic data collected from the connected vehicles traveling along the same route as the ego-vehicle, see~\cite{sun2014velocity,Amini_TRB20,Amini_TCST_2020}.
\end{itemize}
This concept is illustrated in Fig.~\ref{fig:future_information}, where the moving green window represents the short-range high-accuracy speed preview and the orange shrinking windows show the approximate long-term prediction of the future vehicle speed. 

In order to minimize the required computation time of the MH-MPC with a long prediction horizon, the prediction horizon is sampled at different rates. Over the short moving-horizon, a small sampling period of $\delta t_1=1~sec$ is applied and the vehicle speed is updated every $1~sec$. Over the longer shrinking horizon, a large sampling period of $\delta t_2=20~sec$ is used. Given the two horizons, the baseline MPC formulation is revised as follows for the MH-MPC:

\begin{equation}\label{eq:variable_timescale_MPC_formulation}
\begin{split}
\text{min}~\ell_{MH} = \text{min}
\{\sum_{i=t}^{t+N-1}{\dot{m}}_{fuel}\big(P_{eng}(i),T_{cl}(i)\big)\delta t_1~~\\
+\sum_{j=t+N}^{t_{end}}{\dot{m}}_{fuel}\big(P_{eng}(j),T_{cl}(j)\big)\delta t_2\}
\end{split}
\end{equation}
subject to the same constraints (\ref{eq:baseline_constraints_1}) and (\ref{eq:baseline_constraints_2}), and \vspace{-0.15cm}
\begin{gather}\label{eq:variable_timescale_boundary}
SOC_{init}=SOC_{end}.
\end{gather}

In~(\ref{eq:variable_timescale_MPC_formulation}), $N$ is the short moving-horizon, $t$ indicates the current time, and $t_{end}$ is the final time at the end of the trip. Note that there is no terminal penalty in the MH-MPC cost ($\ell_{MH}$). Instead, a terminal constraint is imposed on the battery $SOC$ (\ref{eq:variable_timescale_boundary}).~
Enforcing this constraint is hard and often causes infeasibility for the MH-MPC optimization problem. Thus, this hard constraint is slightly relaxed by allowing the final $SOC$ to vary by $\pm1\%$ as compared with $SOC_{init}$:\vspace{-0.15cm}
\begin{gather}\label{eq:relaxed_condition}
0.99\times SOC_{init} \le SOC_{end} \le SOC_{init}\times1.01.
\end{gather}

Once the MH-MPC optimization problem is solved, the computed control input at the current time ($t$) is applied to the system and the receding horizon is shifted by $1~sec$. 

\vspace{-0.1cm} 
\section{MH-MPC Simulation Results and Discussion}\vspace{-0.1cm} 
\label{sec:sec_4}
\subsection{Simulation Results of MH-MPC}
%
This section presents the simulation results of the proposed MH-MPC over the NYCC and the truncated NEDC. The MH-MPC results are also compared with the DP, the baseline MPC, and a rule-based power-split logic to show the effectiveness of the proposed controller. The rule-based power-split controller is based on the load-leveling logic presented in~\cite{liu2005modeling}.
~Additionally, for the coolant temperature regulation, an extra criteria is included in the rule-based controller to ensure $T_{cl}\geq 50^oC$~\cite{kim2014thermal}. Note that the lower hard constraint imposed on $T_{cl}$ is set to $40^oC$ in the baseline MPC and MH-MPC with access to the speed preview. In the absence of speed preview, a relatively higher temperature (e.g., $50^oC$) is often selected~\cite{kim2014thermal}. Furthermore, the moving-horizon of the MH-MPC is assumed to be $N=20~(20~sec)$ with $\delta t_1=1~sec$. 
\vspace{-0.35cm}
\begin{figure}[h!]
	\begin{center}
		\includegraphics[width=0.99\columnwidth]{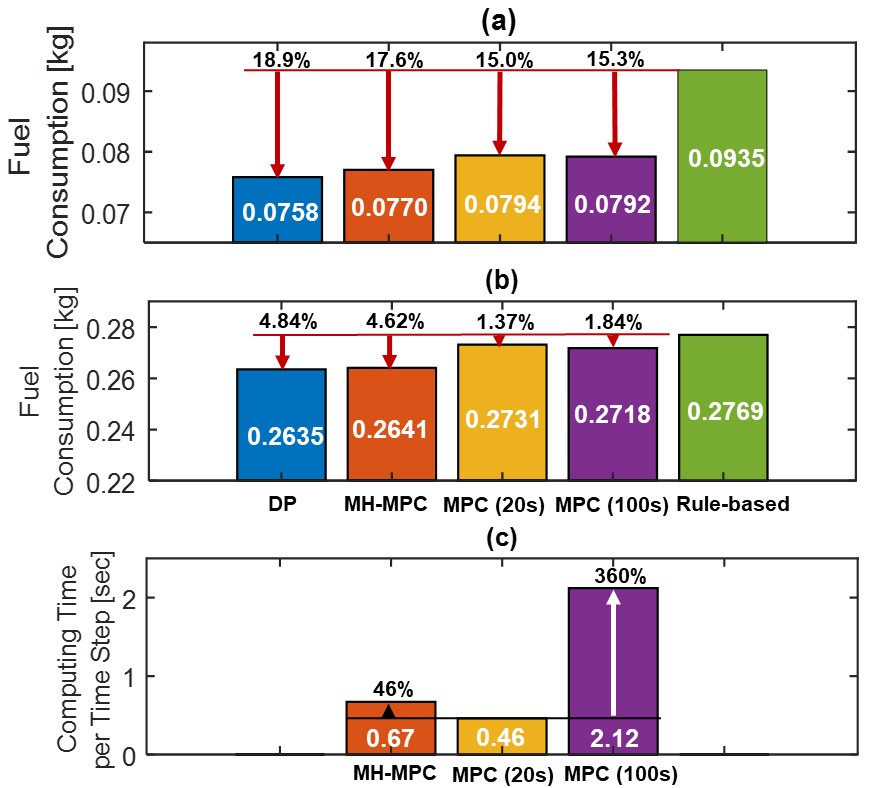}
		\vspace{-0.8cm}   
		\caption{Fuel consumption results of applying different MPCs, DP, and rule-based controller over (a) NYCC (b) truncated NEDC. The average MPC computation time per step time is shown in subplot (c) for baseline MPC with $H=20~(20~sec)$, $100~(100~sec)$, and MH-MPC with $N=20~(20~sec)$.}
		\vspace{-0.6cm} \label{fig:Fuel_Variable} 
	\end{center}
\end{figure}

The comparison of fuel consumption among different methods is shown in Fig.~\ref{fig:Fuel_Variable}. By comparing the rule-based and DP controllers, it can be seen that the DP can reduce the fuel consumption by 19.9\% (NYCC) and 4.8\% (NEDC). These improvements are considered as the maximum achievable fuel saving when the entire driving cycle is known a \textit{priori}. 
While the baseline MPC shows up to 15.3\% fuel saving over NYCC, as compared to the rule-based, the MH-MPC provides further fuel saving of 17.6\% (Fig.~\ref{fig:Fuel_Variable}-($a$)). For the truncated NEDC, the baseline MPC provides marginal fuel saving of $1.37\%$ ($H=20$) and $1.84\%$ ($H=100$) compared to the rule-based controller as shown in Fig.~\ref{fig:Fuel_Variable}-($b$). The MH-MPC, on the other hand, delivers fuel saving of $4.62\%$ over truncated NEDC, which is close to the maximum fuel saving resulted from DP.

The average computation time of the MH-MPC per optimization iteration is recorded and compared against the baseline MPC in Fig.~\ref{fig:Fuel_Variable}-($c$). 
~For MH-MPC, the average computation time was recorded at $0.67~sec$. Comparing with baseline MPC with computation times of $0.46~sec$ ($H=20$) and $2.12~sec$ ($H=100$), one can see that the MH-MPC provides a computationally efficient framework thanks to the incorporated low-resolution long shrinking horizon. For a longer driving cycle, sampling time of the shrinking horizon can be further increased to reduce the computational cost.

\subsection{Leveraging ``uncertain'' speed preview for fuel saving}
\begin{figure}[t!]
	\begin{center}
	    \includegraphics[width=0.8\columnwidth]{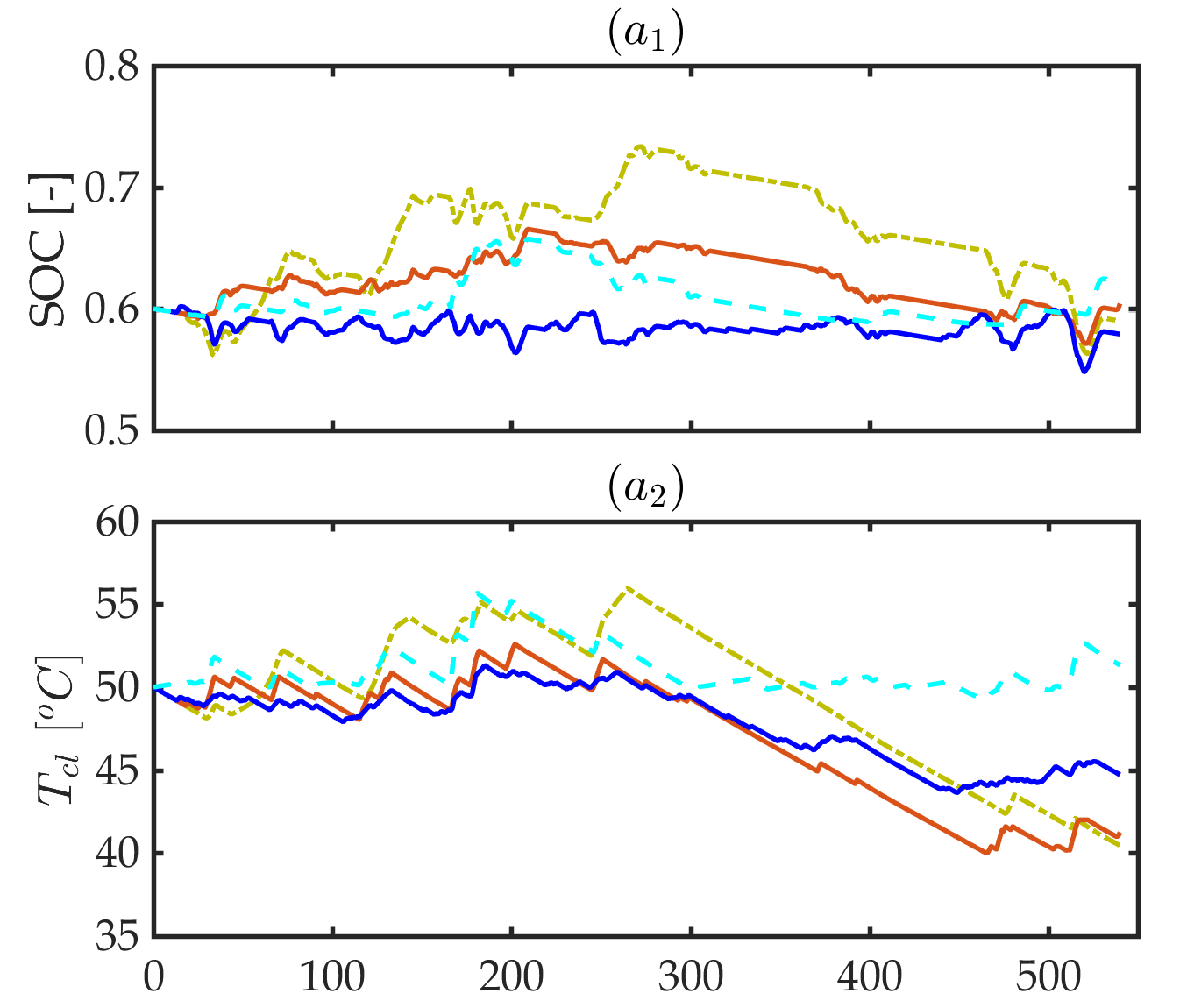}	
		\includegraphics[width=0.8\columnwidth]{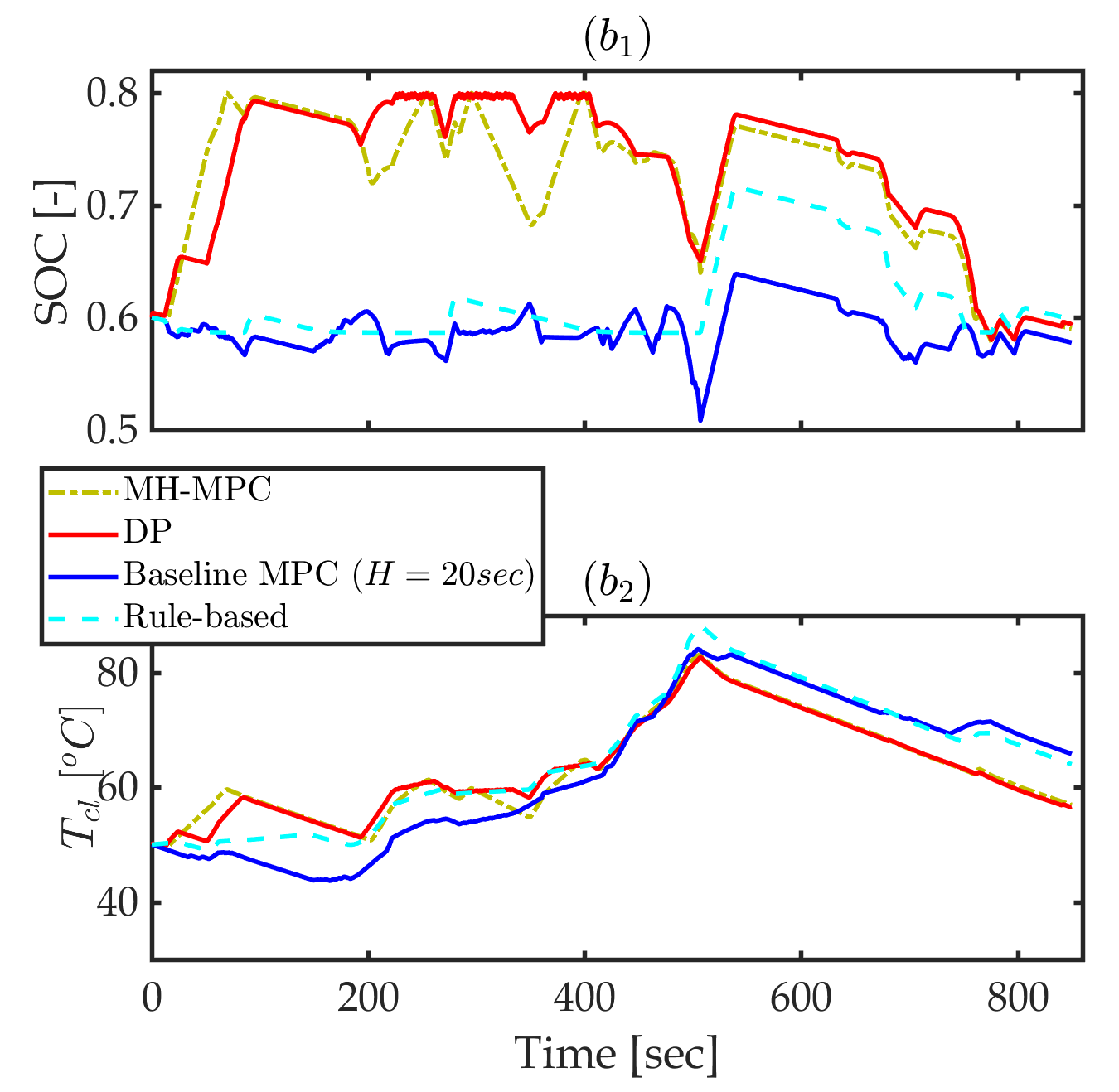}
		\vspace{-0.4cm}   
		\caption{The state trajectories ($SOC$, $T_{cl}$) of the baseline MPC ($H=20$), MH-MPC ($N=20$), DP, and rule-based controllers over (a) NYCC (b) truncated NEDC.}
		\vspace{-0.9cm} \label{fig:Variable_Results} 
	\end{center}
\end{figure}

To explain the superior performance of the MH-MPC, the state trajectories ($SOC$, $T_{cl}$) of different controllers are shown in Fig.~\ref{fig:Variable_Results}. Compared to the baseline MPC and rule-based controller, the $SOC$ trajectories from the MH-MPC vary in a relatively wider range, specifically for the NEDC with $SOC$ trajectory showing a similar trend with DP (Fig.~ \ref{fig:Variable_Results}-($b_1$)). According to the long-range speed preview used in the MH-MPC, the controller is aware of the overall trends along the driving cycle. For instance, along the truncated NEDC, the ego-vehicle drives on the highway for $300~sec$ before it comes to a long stop around $t = 550~sec$. Since the vehicle drives at low speed and with multiple stop-and-go towards the end of the cycle ($t=540-800~sec$), the vehicle can drive in electric vehicle (EV) mode (and save fuel) if enough energy is stored in the battery when it exits the highway. Since the MH-MPC is aware of this upcoming city-driving phase in advance, it commands the battery to be charged while the vehicle drives on the highway (Fig.~ \ref{fig:Variable_Results}-($b_1$)). With the baseline MPC, however, the battery does not have enough charge and the engine is being inefficiently used, see Figs.~\ref{fig:Variable_Results}-($b_2$).
~While the cabin heating is satisfied and the coolant temperature is enforced within its constraints, extra heat is generated and stored in the coolant by the baseline-MPC and rule-based controller, which will be wasted eventually.

\vspace{-0.15cm}
\subsection{Exploiting the Coolant as a Thermal Energy Storage}
For both driving cycles shown in Fig.~\ref{fig:Variable_Results}, it can be seen that during the first part of the trip the $SOC$ and $T_{cl}$ are rising. Then, since the MH-MPC (i) knows the end of the trip, (ii) has charged the battery sufficiently, and (iii)  has stored enough thermal energy in the engine coolant, the vehicle can operate in EV mode while delivering the heating demand to the cabin.
~This favorable response is achieved by exploiting the engine coolant as a energy storage via the developed MH-MPC. The MH-MPC allows for storing the thermal energy in the coolant and releasing it during those periods when traction power demand can be delivered by the battery. 

Note that, while the exact driving cycle is not known a \textit{priori}, MH-MPC incorporates the approximate knowledge about the overall driving cycle trend to shift the thermal loads, thereby enabling the coolant to be leveraged as an energy storage. Additionally, due to the slow thermal dynamics of $T_{cl}$, accurate and second-by-second predictions of the future vehicle speed in long-term is not needed to achieve an effective thermal load shift~\cite{AminiCDC18,Amini_TCST_2019}.

\vspace{-0.15cm}   
\section{Summary and Conclusions}
\label{sec:sec_5}
A multi-horizon MPC (MH-MPC) for integrated power and thermal management (iPTM) of power-split hybrid electric vehicles (HEVs) was developed and studied in this paper. The iPTM was formulated as a constrained nonlinear optimization problem to minimize the overall vehicle fuel consumption while satisfying the traction power and cabin heating demands in a cold-weather condition, and enforcing the constraints on the power ($SOC$) and thermal ($T_{cl}$) dynamics. In order to account for the slow dynamics of the thermal systems and improve the computational efficiency, a multi-horizon prediction horizon was incorporated in the MH-MPC, which includes a short and accurate moving-horizon with a fast update rate, and a long shrinking-horizon with a slower update rate.~
Over the longer shrinking horizon, an approximate estimation of the future vehicle speed is used. The simulation results over urban city driving cycles showed that the proposed MH-MPC provides results similar to those obtained from an offline DP solution with much less computational efforts. Additionally, the MH-MPC allows for exploiting the engine coolant as thermal energy storage. The simulation results showed that $4.6\%-17.6\%$ fuel economy improvement can be achieved over urban driving cycles compared with a conventional rule-based controller.

\vspace{-0.2cm}
\bibliographystyle{unsrt} 
\bibliography{ACC2020Ref.bib}

\begin{thebibliography}{10}

\bibitem{gong2019integrated}
X.~Gong, H.~Wang, M.~Amini, I.~Kolmanovsky, and J.~Sun.
\newblock {Integrated Optimization of Power Split, Engine Thermal Management,
  and Cabin Heating for Hybrid Electric Vehicles}.
\newblock In {\em 3rd CCTA}, 2019.
\newblock {Hong Kong, China}.

\bibitem{kim2016thermal}
N.~Kim and A.~Rousseau.
\newblock {Thermal Impact on the Control and the Efficiency of the 2010 Toyota
  Prius Hybrid Electric Vehicle}.
\newblock {\em Proceedings of the Institution of Mechanical Engineers, Part D:
  Journal of Automobile Engineering}, 230(1):82--92, 2016.

\bibitem{Amini_CCTA19}
M.R. Amini, Y.~Yiheng, H.~Wang, I.~Kolmanovsky, and J.~Sun.
\newblock {Thermal Responses of Connected HEVs Engine and Aftertreatment
  Systems to Eco-Driving}.
\newblock In {\em 3rd CCTA}, 2019.
\newblock {Hong Kong, China}.

\bibitem{kim2014thermal}
N.~Kim, A.~Rousseau, D.~Lee, and H.~Lohse-Busch.
\newblock {Thermal Model Development and Validation for 2010 Toyota Prius}.
\newblock 2014.
\newblock {SAE Technical Paper 2014-01-1784}.

\bibitem{malikopoulos2014supervisory}
A.~Malikopoulos.
\newblock {Supervisory Power Management Control Algorithms for Hybrid Electric
  Vehicles: A Survey}.
\newblock {\em IEEE Transactions on Intelligent Transportation Systems},
  15(5):1869--1885, 2014.

\bibitem{tie2013review}
S.~Tie and C.~Tan.
\newblock {A Review of Energy Sources and Energy Management System in Electric
  Vehicles}.
\newblock {\em Renewable and Sustainable Energy Reviews}, 20:82--102, 2013.

\bibitem{zhang2015comprehensive}
P.~Zhang, F.~Yan, and C.~Du.
\newblock {A Comprehensive Analysis of Energy Management Strategies for Hybrid
  Electric Vehicles based on Bibliometrics}.
\newblock {\em Renewable and Sustainable Energy Reviews}, 48:88--104, 2015.

\bibitem{wei2019optimal}
C.~Wei, T.~Hofman, E.~Caarls, and R.~van Iperen.
\newblock {Optimal Control of an Integrated Energy and Thermal Management
  System for Electrified Powertrains}.
\newblock In {\em ACC}, 2019.
\newblock {Philadelphia, PA, USA}.

\bibitem{shahbakhti2019}
N.~Doshi, D.~Hanover, S.~Hemmati, C.~Morgan, and M.~Shahbakhti.
\newblock {Modeling of Thermal Dynamics of a Connected Hybrid electric Vehicle
  for Integrated HVAC and Powertrain Optimal Operation}.
\newblock In {\em DSCC}, 2019.
\newblock Park City, UT, USA.

\bibitem{shams2012integrated}
M.~Shams-Zahraei, A.~Kouzani, S.~Kutter, and B.~B{\"a}ker.
\newblock {Integrated Thermal and Energy Management of Plug-in Hybrid Electric
  Vehicles}.
\newblock {\em Journal of Power Sources}, 216:237--248, 2012.

\bibitem{lin2003power}
C.~Lin, H.~Peng, J.~Grizzle, and J.~Kang.
\newblock {Power Management Strategy for a Parallel Hybrid Electric Truck}.
\newblock {\em IEEE Transactions on Control Systems Technology},
  11(6):839--849, 2003.

\bibitem{brahma2000opt}
A.~Brahma, Y.~Guezennec, and G.~Rizzoni.
\newblock {Optimal Energy Management in Series Hybrid Electric Vehicles}.
\newblock In {\em ACC}, 2000.
\newblock {Chicago, IL, USA, USA}.

\bibitem{amini2019sequential}
M.R. Amini, X.~Gong, Y.~Feng, H.~Wang, I.~Kolmanovsky, and J.~Sun.
\newblock {Sequential Optimization of Speed, Thermal Load, and Power Split in
  Connected HEVs}.
\newblock In {\em ACC}, 2019.
\newblock {Philadelphia, PA, USA}.

\bibitem{musardo2005ecms}
C.~Musardo, G.~Rizzoni, Y.~Guezennec, and B.~Staccia.
\newblock {A-ECMS: An Adaptive Algorithm for Hybrid Electric Vehicle Energy
  Management}.
\newblock {\em European Journal of Control}, 11(4-5):509--524, 2005.

\bibitem{kim2011optimal}
N.~Kim, S.~Cha, and H.~Peng.
\newblock {Optimal Equivalent Fuel Consumption for Hybrid Electric Vehicles}.
\newblock {\em IEEE Transactions on Control Systems Technology},
  20(3):817--825, 2011.

\bibitem{onori2011adaptive}
S.~Onori, L.~Serrao, and G.~Rizzoni.
\newblock {Adaptive Equivalent Consumption Minimization Strategy for Hybrid
  Electric Vehicles}.
\newblock In {\em DSCC}, 2011.
\newblock {Arlington, VA, USA}.

\bibitem{borhan2011mpc}
H.~Borhan, A.~Vahidi, A.~Phillips, M.~Kuang, I.~Kolmanovsky, and S.~Di~Cairano.
\newblock {MPC-based Energy Management of a Power-Split Hybrid Electric
  Vehicle}.
\newblock {\em IEEE Transactions on Control Systems Technology},
  20(3):593--603, 2011.

\bibitem{zhang2016real}
J.~Zhang and T.~Shen.
\newblock {Real-time Fuel Economy Optimization with Nonlinear MPC for PHEVs}.
\newblock {\em IEEE Transactions on Control Systems Technology},
  24(6):2167--2175, 2016.

\bibitem{wang2016model}
H.~Wang, Y.~Huang, A.~Khajepour, and Q.~Song.
\newblock {Model Predictive Control-based Energy Management Strategy for a
  Series Hybrid Electric Tracked Vehicle}.
\newblock {\em Applied Energy}, 182:105--114, 2016.

\bibitem{yang2019eco}
Z.~Yang, Y.~Feng, X.~Gong, D.~Zhao, and J.~Sun.
\newblock {Eco-Trajectory Planning with Consideration of Queue along Congested
  Corridor for Hybrid Electric Vehicles}.
\newblock {\em Transportation Research Record}, 2019.
\newblock in press, doi:10.1177/0361198119845363.

\bibitem{Amini_TCST_2019}
M.R. Amini, H.~Wang, X.~Gong, D.~Liao-McPherson, I.~Kolmanovsky, and J.~Sun.
\newblock {Cabin and Battery Thermal Management of Connected and Automated HEVs
  for Improved Energy Efficiency Using Hierarchical Model Predictive Control}.
\newblock {\em IEEE Transactions on Control Systems Technology}, 2019.
\newblock in press, doi:10.1109/TCST.2019.2923792.

\bibitem{Amini_TCST_2020}
M.R. Amini, I.~Kolmanovsky, and J.~Sun.
\newblock {Hierarchical MPC for Robust Eco-Cooling of Connected and Automated
  Vehicles and Its Application to Electric Vehicle Battery Thermal Management
  }.
\newblock {\em IEEE Transactions on Control Systems Technology}, 2020.
\newblock in press, doi:10.1109/TCST.2020.2975464.

\bibitem{Amini_TRB20}
M.R. Amini, Y.~Feng, Z.~Yang, I.~Kolmanovsky, and J.~Sun.
\newblock {Long-term Vehicle Speed Prediction via Traffic Data Mining for
  Improved Energy Efficiency of Connected Electric Vehicles}.
\newblock In {\em TRB 99th Annual Meeting}, 2020.
\newblock {Washington, D.C., USA}.

\bibitem{poramapojana2012minimizing}
P.~Poramapojana and B.~Chen.
\newblock {Minimizing HEV Fuel Consumption using Model Predictive Control}.
\newblock In {\em IEEE/ASME 8th IEEE/ASME Int. Conference on Mechatronic and
  Embedded Systems and Applications}, 2012.
\newblock {Suzhou, China}.

\bibitem{risbeck2016mpctools}
M.~Risbeck and J.~Rawlings.
\newblock {MPCTools: Nonlinear Model Predictive Control Tools for CasADi}.
\newblock 2016.
\newblock {[online] Available:
  https://bitbucket.org/rawlings-group/octave-mpctools}.

\bibitem{andersson2019casadi}
J.~Andersson, J.~Gillis, G.~Horn, J.~Rawlings, and M.~Diehl.
\newblock {CasADi: a Software Framework for Nonlinear Optimization and Optimal
  Control}.
\newblock {\em Mathematical Programming Computation}, 11(1):1--36, 2019.

\bibitem{AminiCDC18}
M.R. Amini, I.~Kolmanovsky, and J.~Sun.
\newblock {Two-Layer Model Predictive Battery Thermal and Energy Management
  Optimization for Connected and Automated Electric Vehicles}.
\newblock In {\em 57th CDC}, 2018.
\newblock {Miami Beach, FL, USA}.

\bibitem{Amini_CDC19}
M.R. Amini, I.~Kolmanovsky, and J.~Sun.
\newblock {Robust Hierarchical MPC for Handling Long Horizon Demand Forecast
  Uncertainty with Application to Automotive Thermal Management}.
\newblock In {\em 58th CDC}, 2019.
\newblock {Nice, France}.

\bibitem{sun2014velocity}
C.~Sun, X.~Hu, S.~Moura, and F.~Sun.
\newblock {Velocity Predictors for Predictive Energy Management in Hybrid
  Electric Vehicles}.
\newblock {\em IEEE Transactions on Control Systems Technology},
  23(3):1197--1204, 2014.

\bibitem{liu2005modeling}
J.~Liu, H.~Peng, and Z.~Filipi.
\newblock {Modeling and Analysis of the Toyota Hybrid System}.
\newblock In {\em Int. Conference on Advanced Intelligent Mechatronics}, 2005.
\newblock {Monterey, CA, USA}.

\end{thebibliography}

\end{document}